\documentclass[onecolumn]{IEEEtran}

\usepackage{setspace}
\ifCLASSINFOpdf
\else
   \usepackage[dvips]{graphicx}
\fi
\usepackage{url}
\usepackage{lipsum}
\usepackage{bm}
\usepackage{graphicx}
\usepackage{amsmath,amsfonts,amssymb,amsthm}

\usepackage{xcolor}
\usepackage{anyfontsize}
\usepackage{mathtools} 
\usepackage{algorithm}
\usepackage{algorithmicx}
\usepackage{algpseudocode}
\usepackage{comment}
\usepackage{titlesec}

\usepackage{graphicx}
\usepackage{subfigure}  
\usepackage{booktabs}

\begin{document}

\title{Stochastic Expectation Maximization for Robust State-Space Radio Interferometric Imaging}

\author{Nawel Arab$^{\dagger}$,  Mohammed Nabil El Korso$^{\ddagger}$, Isabelle Vin$^{\dagger}$  and Pascal Larzabal$^{\dagger}$\\

\vspace{1em}
 $^{\dagger}$SATIE, ENS Paris-Saclay, Université Paris-Saclay, France\\
 $^{\ddagger}$L2S, CentraleSupelec, Université Paris-Saclay, France\\

}
\maketitle

\begin{abstract}
State--space models provide a flexible framework for analyzing dynamical systems, yet
they often rely on Gaussian assumptions that fail to capture heavy-tailed or
outlier-prone measurement noise. We propose a robust estimation scheme for linear
state--space models subject to compound-Gaussian noise, as encountered for instance in
radio interferometry affected by radio-frequency interference (RFI). The
method relies on a Stochastic Approximation Expectation--Maximization (SAEM) algorithm
in which the standard E-step is replaced by Monte Carlo sampling of the latent states
and noise texture through closed-form Gibbs updates, enabling tractable inference
despite the heavy-tailed likelihood. Numerical experiments show that the proposed method
significantly improves reconstruction fidelity and robustness to RFI, outperforming a
Gaussian EM algorithm and even an oracle RTS smoother. These results highlight the
benefits of heavy-tailed state--space modeling and SAEM-based inference in
interference-dominated imaging scenarios.
\end{abstract}


\IEEEpeerreviewmaketitle

\section{Introduction}

State-space models provide a powerful framework for describing the evolution of hidden
states in dynamical systems \cite{b1,b2,Aravkin2014}. Conventionally, state-space models assume
Gaussian measurement and state noise, owing to their tractability and well-characterized
statistical properties. However, many real-world phenomena are subject to perturbations
that deviate from the conventional Gaussian noise assumption. In radio interferometry,
for instance, observational data are frequently corrupted by non-Gaussian noise sources
such as radio-frequency interference (RFI) \cite{leshem,Kazemi2013}, which originates from
man-made signals and introduces significant distortions into astronomical measurements
\cite{leshem2,nawel}. Such interference produces sporadic high-power spikes in the
measured visibilities, leading to heavy-tailed statistics. Many radio-interferometric reconstruction methods assume Gaussian additive noise
\cite{wiaux,carrillo,onose,pratley}, an approximation that can lead to
inaccurate reconstructions when the heavy-tailed nature of real-world measurement
noise is not properly accounted for.

In the realm of state-space modeling, addressing non-Gaussian noise has led to the
development of various methodological approaches, notably particle filtering and
non-conventional Kalman filters. Particle filters \cite{doucet}, or Sequential Monte
Carlo methods, are designed to handle non-linear and non-Gaussian state-space models by
representing the posterior distribution with a set of weighted samples
\cite{pf1,pf2,abdallah2020bayesian}. However, they present several weaknesses, such as particle degeneracy
and high computational cost, which are exacerbated in high-dimensional systems where
the number of particles required for accurate approximation grows rapidly. Robust
filters and smoothers, on the other hand, often based on variational Bayes or
generalized maximum-likelihood principles, replace the Gaussian measurement model with
heavy-tailed distributions such as the Student's $t$ or Laplace laws \cite{k2,k3,k4,zhang2016maximum}.
While these filters can attenuate the influence of outliers, a critical and often
unaddressed limitation is that they typically assume the model parameters, such as
noise covariances, are known \emph{a priori} \cite{k5}.

This gap in the literature motivates the present work: we address the problem of joint
state and parameter estimation in linear SSMs corrupted by heavy-tailed noise. We adopt
a hierarchical Bayesian formulation where the measurement noise is modeled as a
compound-Gaussian distribution \cite{delmas,hippert2025missing}, which marginalizes to a Student's $t$
likelihood. This representation introduces latent scale (or texture) variables which,
when conditioned upon, restore the conditional linearity and Gaussianity of the model.
While this structure is theoretically appealing, the joint posterior over states,
scales, and parameters becomes analytically intractable.

To overcome this intractability, we propose a Stochastic Approximation
Expectation–Maximization (SAEM) algorithm \cite{sem,sem2}. SAEM extends the classical
EM framework of \cite{dempster} by replacing the intractable expectation (E-step) with
a stochastic approximation based on Monte Carlo samples. The key innovation of our work
lies in the design of an efficient simulation step: we develop a block Gibbs sampler
that exploits the model's conditional structure to sample jointly from the latent
states and scales. Specifically, the sampler alternates between (i) sampling the entire
state trajectory in a single block using a Forward–Filtering Backward–Sampling (FFBS)
procedure, made possible by the conditionally Gaussian structure, and (ii) sampling the
latent scale variables independently from their Gamma full conditional distributions.
This approach combines the stability and convergence properties of EM with the
flexibility of Monte Carlo methods. Furthermore, we carefully derive the 
transformations to handle complex-valued observations, such as radio-interferometric
visibilities, within a real-valued Kalman filtering framework, ensuring computational
practicality.

We validate the efficacy of the robust SAEM algorithm on a synthetic dynamic
radio-interferometric imaging scenario, simulating a rotating ring source observed by
the Very Large Array (VLA) configuration and contaminated by realistic RFI. Our
experimental results demonstrate that the proposed method achieves a substantial
improvement in reconstruction fidelity, over  strong baselines: a standard Gaussian EM algorithm and an oracle RTS smoother that has access to the true model parameters.

\section{Problem formulation}
We consider a discrete-time linear state-space model with compound-Gaussian additive measurement noise
\begin{align}
    \mathbf{x}_k &= \mathbf{F}_k \mathbf{x}_{k-1} + \mathbf{w}_k, 
    & \mathbf{w}_k &\sim \mathcal{N}(\mathbf{0},\mathbf{Q}_k), 
    \label{eq:state_transition}\\[2mm]
    \mathbf{y}_k &= \mathbf{H}_k \mathbf{x}_k + \mathbf{\Omega}_k^{-1/2}\,\mathbf{n}_k, 
    & \mathbf{n}_k &\sim \mathcal{CN}\!\bigl(\mathbf{0},\mathbf{R}_k\bigr),
    \label{eq:observation_model}
\end{align}
where $\mathbf{x}_k \in \mathbb{R}^n$ is the latent state, $\mathbf{y}_k \in \mathbb{C}^m$ is the observation vector, $\mathbf{F}_k \in \mathbb{R}^{n\times n}$ is the state transition matrix, and $\mathbf{H}_k \in \mathbb{C}^{m\times n}$ is the measurement matrix. The process noise covariance is $\mathbf{Q}_k \in \mathbb{R}^{n\times n}$, and $\mathbf{R}_k \in \mathbb{C}^{m\times m}$ is a diagonal Hermitian positive-definite matrix specifying the covariance of the complex noise $\mathbf{n}_k$. The initial state \(\mathbf{x}_0 \sim \mathcal{N}(\bm{\mu}_0,\bm{\Sigma}_0)\), process noise, and measurement noise are assumed to be mutually independent.

The observation noise is modeled as a compound Gaussian distribution, 
defined as the product of two independent random variables: a complex Gaussian speckle $\mathbf{n}_k$ and a positive random texture matrix 
\(\mathbf{\Omega}_k = \mathrm{diag}(\tau_{k,1},\ldots,\tau_{k,m}) \),
whose diagonal entries are i.i.d. Gamma random variables 
\begin{equation*}
    \tau_{k,i} \sim \text{Gamma}\!\left(\frac{\nu}{2}, \frac{\nu}{2}\right), 
    \qquad i=1,\ldots,m, \quad \nu>2,
\end{equation*}
following the shape-rate parametrization so that $\mathbb{E}[ \tau_{k,i}] =1$. Conditional on $\mathbf{\Omega}_k$, the measurement noise is Gaussian,

\begin{equation}
       \mathbf{y}_k \mid \mathbf{x}_k,\mathbf{\Omega}_k \sim 
    \mathcal{CN}\!\bigl(\mathbf{H}_k \mathbf{x}_k,\,
     \mathbf{\Omega}_k^{-1}\mathbf{R}_k\bigr). 
\end{equation}


Marginalizing out the texture variables yields a heavy-tailed likelihood for the observations. In particular, under the above Gamma prior and independence assumptions between $\mathbf{n}_k$ and $\mathbf{\Omega}_k$, the conditional density of $\mathbf{y}_k$ given $\mathbf{x}_k$ is a component-wise complex Student's $t$-distribution, whose joint probability density function can be written as

{\footnotesize
\begin{equation}
      p(\mathbf{y}_k \mid \mathbf{x}_k) = \frac{\Gamma(m+n)}{\Gamma(\nu)(\pi\nu)^m} \left[ 1 + \frac{1}{\nu}(\mathbf{y}_k - \mathbf{H}_k\mathbf{x}_k)^\dag \mathbf{R}_k^\dag (\mathbf{y}_k - \mathbf{H}_k\mathbf{x}_k)\right]^{-(\nu+m)},\label{eq:student}
\end{equation}}
where $(\cdot)^{\dagger}$ denotes the Hermitian transpose. This heavy-tailed observation model provides robustness to outliers and better matches the non-Gaussian interferences and clutters encountered in many signal processing applications~\cite{finegold,liu,meriaux2020matched}. 


Given a sequence of observations $\mathbf{y}_{1:K} = \{\mathbf{y}_1,\ldots,\mathbf{y}_K\}$, we aim to jointly reconstruct the hidden state trajectory $\mathbf{x}_{0:K}$ and the model parameters $\bm{\theta} = \{\bm{\mu}_0,\bm{\Sigma}_0,\{\mathbf{Q}_k\}_{k=1}^K,\{\mathbf{R}_k\}_{k=1}^K\}$. To achieve this, we adopt the complete-data formulation, where we treat both the states $\mathbf{x}_{0:K}$ and the auxiliary scale matrices $\mathbf{\Omega}_{1:K}$ as latent variables. For a fixed
$\bm{\theta}$, the complete-data joint density factorizes according to the
state--space structure as
{{\fontsize{8.5}{12}
\begin{equation}
    p_{\bm{\theta}}\bigl(\mathbf{x}_{0:K},\mathbf{\Omega}_{1:K}, \mathbf{y}_{1:K}\bigr)
    =
    p_{\bm{\theta}}(\mathbf{x}_0) \prod_{k=1}^{K} p_{\bm{\theta}}(\mathbf{y}_{k}|  \mathbf{x}_{k}, \mathbf{\Omega}_k) p_{\bm{\theta}}(\mathbf{x}_{k}|\mathbf{x}_{k-1}) p( \bm{\Omega}_{k}),
    \label{eq:complete_joint}
\end{equation}
}
where $p_{\bm{\theta}}(\mathbf{y}_k \mid \mathbf{x}_k,\mathbf{\Omega}_k)$ is the conditionally Gaussian likelihood and $p(\mathbf{\Omega}_k)$ denotes the Gamma prior induced/associated by the compound Gaussian construction.

The complete-data formulation~\eqref{eq:complete_joint} provides a convenient starting
point for statistical inference, since the log-likelihood decomposes into
terms that depend on the states, the textures and the parameters in a structured way. However, recovering the latent trajectory
$\mathbf{x}_{0:K}$ and the texture variables $\mathbf{\Omega}_{1:K}$ from the
observations is challenging. The heavy-tailed measurement model induces a strong
coupling between states and textures once conditioned on the data, making exact
inference analytically intractable. Several approaches have been proposed for such
models, including particle filtering, particle MCMC, and variational approximations,
each trading accuracy for computational cost and assuming $\bm\theta$ is known. In this work, we
adopt an EM inference strategy, which naturally leverages the
complete-data joint structure, and yields stable and closed form parameter updates.

\section{Expectation--Maximization Framework}

We cast $(\mathbf{x}_{0:K},\mathbf{\Omega}_{1:K})$ as latent variables and adopt the
EM algorithm to jointly estimate the latent states and the
parameters. Let
\(\bm{\xi} = \{\mathbf{y}_{1:K},\mathbf{x}_{0:K},\mathbf{\Omega}_{1:K}\} \) denote the complete data. Starting from an initial parameter value
$\bm{\theta}^{(0)}$, the EM constructs a sequence $\{\bm{\theta}^{(i)}\}_{i\geq 0}$ by
alternating the following two steps:

\emph{E-step.} Given $\bm{\theta}^{(i)}$, compute the conditional expectation of the
complete-data log-likelihood,
\begin{equation}
    \mathcal{Q}(\bm{\theta} \mid \bm{\theta}^{(i)})
    =
    \mathbb{E}_{\mathbf{x}_{0:K},\mathbf{\Omega}_{1:K} \mid
    \mathbf{y}_{1:K},\bm{\theta}^{(i)}}
    \Big[
        \log p_{\bm{\theta}}(\mathbf{y}_{1:K},\mathbf{x}_{0:K},\mathbf{\Omega}_{1:K})
    \Big],
    \label{eq:Q_function}
\end{equation}
where the expectation is taken with respect to the conditional distribution 
$p_{\bm{\theta}^{(i)}}(\mathbf{x}_{0:K},\mathbf{\Omega}_{1:K} | \mathbf{y}_{1:K})$ induced by the current parameter estimate
$\bm{\theta}^{(i)}$.

\emph{M-step.} Update the parameters by maximizing the surrogate function,
\begin{equation}
    \bm{\theta}^{(i+1)} \in
    \arg\max_{\bm{\theta}} \,
    \mathcal{Q}(\bm{\theta} \mid \bm{\theta}^{(i)}).
    \label{eq:M_step}
\end{equation}

In the present robust state--space model, exact computation of the
expectation in~\eqref{eq:Q_function} cannot be carried out in closed form. Although the states
$\mathbf{x}_k$ and the texture matrices $\mathbf{\Omega}_k$ are a priori independent,
conditioning on the observations $\mathbf{y}_{1:K}$ induces a strong coupling through the heavy-tailed likelihood. As a result, the joint posterior $p_{\bm{\theta}^{(i)}}(\mathbf{x}_{0:K},\mathbf{\Omega}_{1:K} \mid \mathbf{y}_{1:K})$ is intractable and $\mathcal{Q}(\bm{\theta} \mid \bm{\theta}^{(i)})$ does
not admit a closed-form expression.

To address this limitation, we rely on a stochastic approximation EM (SAEM)
scheme in which the expectation in~\eqref{eq:Q_function} is approximated by Monte
Carlo samples \cite{sem2}. More precisely, we exploit the state--space factorization of the complete-data joint
in~\eqref{eq:complete_joint} to derive closed-form full conditional distributions for
$\mathbf{x}_{0:K}$ and $\mathbf{\Omega}_{1:K}$. This yields a block Gibbs sampler
targeting the joint posterior
$p_{\bm{\theta}^{(i)}}(\mathbf{x}_{0:K},\mathbf{\Omega}_{1:K} \mid \mathbf{y}_{1:K})$,
and the resulting samples are then used to construct a stochastic approximation of the
auxiliary $\mathcal{Q}$-function.

\subsection{Stochastic Approximation }


SAEM replaces the exact auxiliary $\mathcal{Q}$-function
 by a recursively updated stochastic
approximation. At iteration $i$, given the current parameter $\bm{\theta}^{(i)}$, a new sample of the
latent variables
\begin{equation}
    \bigl(\mathbf{x}_{0:K}^{(i+1)},\mathbf{\Omega}_{1:K}^{(i+1)}\bigr)
    \sim p_{\bm{\theta}^{(i)}}(\mathbf{x}_{0:K},\mathbf{\Omega}_{1:K}
    \mid \mathbf{y}_{1:K}),\label{eq:sample}
\end{equation}
is generated using a block Gibbs sampler. We then update an estimate
$\widehat{\mathcal{Q}}_i(\bm{\theta})$ of the EM auxiliary function by a
Robbins--Monro recursion of the form
\begin{equation}
    \widehat{\mathcal{Q}}_{i+1}(\bm{\theta})
    =
    \widehat{\mathcal{Q}}_{i}(\bm{\theta})
    + \gamma_{i+1}
    \Big[
        \log p_{\bm{\theta}}(\mathbf{y}_{1:K},
        \mathbf{x}_{0:K}^{(i+1)},\mathbf{\Omega}_{1:K}^{(i+1)})
        - \widehat{\mathcal{Q}}_{i}(\bm{\theta})
    \Big],
    \label{eq:saem_update}
\end{equation}
where $\{\gamma_i\}$ is a decreasing sequence of step sizes. In practice, the recursion is applied to the sufficient statistics of the complete-data model, which leads to simple closed-form parameter updates in the subsequent M-step. 

For the compound-Gaussian
state–space model considered here, a convenient set of sufficient statistics  \(S\bigl(\mathbf{x}_{0:K},\mathbf{\Omega}_{1:K}\bigr)
            = \{S_{xx},S_{xx^-},S_{x^-x^-},S_{xx}^{\tau},S_{yx}^{\tau},S_{yy}^{\tau}\}\) is
\begin{align}
S_{xx}        &= \sum_{k=0}^{K} \mathbf{x}_k \mathbf{x}_k^{\top}, &
S_{xx^{-}}    &= \sum_{k=1}^{K} \mathbf{x}_k \mathbf{x}_{k-1}^{\top}, \nonumber\\
S_{x^- x^-}   &= \sum_{k=1}^{K} \mathbf{x}_{k-1} \mathbf{x}_{k-1}^{\top}, &
S_{xx}^{\tau} &= \sum_{k=0}^{K} \mathbf{\Omega}_k \mathbf{x}_k \mathbf{x}_k^{\top}, \nonumber\\
S_{yx}^{\tau} &= \sum_{k=0}^{K} \mathbf{\Omega}_k \mathbf{y}_k \mathbf{x}_k^{\top}, &
S_{yy}^{\tau} &= \sum_{k=0}^{K} \mathbf{\Omega}_k \mathbf{y}_k \mathbf{y}_k^{\dagger}.\label{eq:sufficient_stats}
\end{align}

SAEM maintains running stochastic approximations of these statistics via
\eqref{eq:saem_update}, and the subsequent M-step then reduces to simple closed-form
updates of $\bm{\theta}$ obtained by replacing empirical sums in the complete-data
log-likelihood with their SAEM-averaged counterparts.

Given the current approximation $\widehat{\mathcal{Q}}_{i+1}(\bm{\theta})$, the
parameter update is obtained by an approximate M-step,
\begin{equation}
    \bm{\theta}^{(i+1)}
    \in
    \arg\max_{\bm{\theta}}
    \,\widehat{\mathcal{Q}}_{i+1}(\bm{\theta}).
    \label{eq:saem_Mstep}
\end{equation}
Thus, each iteration consists of a simulation step \eqref{eq:sample}, a stochastic approximation step
\eqref{eq:saem_update}, and a maximization step \eqref{eq:saem_Mstep}. This procedure
preserves the EM interpretation in terms of complete-data likelihood, while allowing
the use of a small number of Monte Carlo samples per iteration and ensuring
convergence under mild regularity conditions  \cite{sem2}.

\section{Block Gibbs sampler}

At iteration $i$ of the SAEM algorithm, given the current parameter vector
$\bm{\theta}^{(i)}$ and a realization $\mathbf{\Omega}_{1:K}^{(i)}$, a new
sample of the latent variables
$(\mathbf{x}_{0:K}^{(i+1)},\mathbf{\Omega}_{1:K}^{(i+1)})$ is generated by alternating between
sampling the state trajectory and the texture matrices from their full conditional
distributions.

\vspace{0.3em}
\paragraph*{Conditional update of the state trajectory (complex-to-real representation).}
Recall that the states $\mathbf{x}_k\in\mathbb{R}^n$ are real, whereas the measurements
$\mathbf{y}_k\in\mathbb{C}^m$ are complex-valued with
\[
    \mathbf{y}_k \mid \mathbf{x}_k,\mathbf{\Omega}_k
    \sim
    \mathcal{CN}\!\bigl(
        \mathbf{H}_k \mathbf{x}_k,\,
        \mathbf{\Omega}_k^{-1}\mathbf{R}_k
    \bigr).
\]
To work within a real-valued state--space framework, we adopt the real-augmented
representation of circular complex Gaussian vectors,
\[
    \mathbf{y}_k^{\mathbb{R}}
    =
    \begin{bmatrix}
        \Re(\mathbf{y}_k)\\[1mm]
        \Im(\mathbf{y}_k)
    \end{bmatrix}\in\mathbb{R}^{2m},
    \qquad
    \mathbf{H}_k^{\mathbb{R}}
    =
    \begin{bmatrix}
        \Re(\mathbf{H}_k)\\[1mm]
        \Im(\mathbf{H}_k)
    \end{bmatrix} \in\mathbb{R}^{2m\times n},
\]
with real covariance
\[
    \mathbf{\Sigma}_k^{\mathbb{R}}(\mathbf{\Omega}_k)
    =
    \begin{bmatrix}
        \Re\!\big(\mathbf{\Omega}_k^{-1}\mathbf{R}_k\big)
        &
        -\Im\!\big(\mathbf{\Omega}_k^{-1}\mathbf{R}_k\big)
        \\[1mm]
        \Im\!\big(\mathbf{\Omega}_k^{-1}\mathbf{R}_k\big)
        &
        \ \Re\!\big(\mathbf{\Omega}_k^{-1}\mathbf{R}_k\big)
    \end{bmatrix}\in\mathbb{R}^{2m\times 2m}.
\]
The standard mapping between circular complex Gaussians and real-augmented
representations yields the equivalent likelihood
\[
    \mathbf{y}_k^{\mathbb{R}}
    \mid
    \mathbf{x}_k,\mathbf{\Omega}_k
    \sim
    \mathcal{N}\!\left(
        \mathbf{H}_k^{\mathbb{R}} \mathbf{x}_k,\;
        \tfrac{1}{2}\,\mathbf{\Sigma}_k^{\mathbb{R}}(\mathbf{\Omega}_k)
    \right).
\]
In the homoscedastic case $\mathbf{R}_k=\mathbf{I}_m$, this simplifies to
\(
    \mathbf{y}_k^{\mathbb{R}}
    \sim
    \mathcal{N}\!\left(
        \mathbf{H}_k^{\mathbb{R}}\mathbf{x}_k,\;
        \tfrac12\,\mathbf{\Omega}_k^{-1}\otimes\mathbf{I}_2
    \right),
\)
yielding a fully real-valued linear--Gaussian observation model. Conditionally on
$\mathbf{\Omega}_{1:K}$, the posterior of $\mathbf{x}_{0:K}$ is therefore Gaussian, and
we sample $\mathbf{x}_{0:K}$ from
$p_{\bm{\theta}^{(i)}}(\mathbf{x}_{0:K}\mid\mathbf{y}_{1:K},\mathbf{\Omega}_{1:K}^{(i)})$
using standard Kalman filtering followed by a backward simulation step (FFBS).

\vspace{0.3em}
\paragraph*{Conditional update of the scale variables.}
Conversely, conditionally on the latent states $\mathbf{x}_{0:K}$ and the observations
$\mathbf{y}_{1:K}$, the scale variables
$\mathbf{\Omega}_k = \mathrm{diag}(\tau_{k,1},\ldots,\tau_{k,m})$ are independent
across time and measurement components. Let $\mathbf{h}_{k,i}^\dagger$ denote the
$i$th row of $\mathbf{H}_k$. From the hierarchical model,
\begin{equation}
    y_{k,i}\mid \mathbf{x}_k,\tau_{k,i}
    \sim
    \mathcal{CN}\!\left(
        \mathbf{h}_{k,i}^\dagger \mathbf{x}_k,\;
        [\mathbf{R}_k]_{ii}\,\tau_{k,i}^{-1}
    \right),
\end{equation}
with $\tau_{k,i}\sim\mathrm{Gamma}\!\left(\frac{\nu}{2},\frac{\nu}{2}\right)$. The full conditional distribution is obtained by combining the prior and likelihood densities.
Introducing the normalized squared residual
\[
    \delta_{k,i}
    = \frac{\bigl|\,y_{k,i}-\mathbf{h}_{k,i}^\dagger\mathbf{x}_k\,\bigr|^2}
           {[\mathbf{R}_k]_{ii}},
\]
we obtain, in the Gamma shape--rate parameterization,
\begin{equation}
    \tau_{k,i}\mid \mathbf{x}_k,\mathbf{y}_k
    \sim
    \mathrm{Gamma}\!\left(
        \frac{\nu+1}{2},\;
        \frac{\nu+\delta_{k,i}}{2}
    \right),
    \label{eq:tau_posterior}
\end{equation}
for all $i=1,\ldots,m$ and $k=1,\ldots,K$. Thus all texture coefficients $\{\tau_{k,i}\}$
admit closed-form, independent updates. Sampling from~\eqref{eq:tau_posterior} yields a
new realization $\mathbf{\Omega}_{1:K}^{(i+1)}$ in the Gibbs cycle.

\vspace{0.3em}
\paragraph*{Block Gibbs kernel}
Alternating the FFBS step for $\mathbf{x}_{0:K}$ with the Gamma updates for
$\mathbf{\Omega}_{1:K}$ defines a block Gibbs sampler targeting the conditional
distribution of the latent variables.

\paragraph*{Summary of the block Gibbs step :}
For a fixed parameter value $\bm{\theta}^{(i)}$, one full iteration of the block Gibbs sampler alternates:

\begin{itemize}
\item[\emph{(i)}] a Gamma update of all scale variables
      $\mathbf{\Omega}_{1:K}^{(i+1)}$ using~\eqref{eq:tau_posterior}, and
\item[\emph{(ii)}] an FFBS draw of the state trajectory
      $\mathbf{x}_{0:K}^{(i+1)}$ based on the real-augmented linear--Gaussian model.
\end{itemize}

This Markov transition leaves
$p_{\bm{\theta}^{(i)}}(\mathbf{x}_{0:K},\mathbf{\Omega}_{1:K}\mid \mathbf{y}_{1:K})$
invariant and constitutes the simulation step of the SAEM algorithm.


\vspace{0.3em}

To summarize the overall inference procedure, a compact algorithmic description of the
proposed SAEM–Gibbs method is provided in Algorithm~\ref{alg:saem_gibbs}.

\begin{algorithm}
\caption{Robust SAEM--Gibbs algorithm}
\label{alg:saem_gibbs}
\begin{algorithmic}[1]

    \State \textbf{Inputs:} Observations $\mathbf{y}_{1:K}$, model matrices $\{\mathbf{F}_k,\mathbf{H}_k\}$, degrees of freedom $\nu$, step-size sequence $(\gamma_i)$.
    \State \textbf{Initialization:} 
    $\bm{\theta}^{(0)}$, $(\mathbf{x}_{0:K}^{(0)},\bm{\Omega}_{1:K}^{(0)})$, $\widehat{S}^{(0)}$

    \For{$i = 0$ to $I-1$}
        \State \textbf{Simulation (block Gibbs):}
        \State \quad $\mathbf{x}_{0:K}^{(i+1)} \sim
        p_{\bm{\theta}^{(i)}}(\mathbf{x}_{0:K} \mid \mathbf{y}_{1:K},\bm{\Omega}_{1:K}^{(i)})$ \hfill (FFBS)
        \State \quad $\bm{\Omega}_{1:K}^{(i+1)} \sim
        p_{\bm{\theta}^{(i)}}(\bm{\Omega}_{1:K} \mid \mathbf{x}_{0:K}^{(i+1)},\mathbf{y}_{1:K})$ \hfill (see~\eqref{eq:tau_posterior})
        
        \State \textbf{Stochastic approximation step:}
        \State \quad Compute  \(S\bigl(\mathbf{x}_{0:K}^{(i+1)},\mathbf{\Omega}_{1:K}^{(i+1)}\bigr)\) as  in~\eqref{eq:sufficient_stats}.
        \State \quad Update the sufficient statistics according to
        \[
            \widehat{S}^{(i+1)}
            =
            (1-\gamma_{i+1})\,\widehat{S}^{(i)}
            + \gamma_{i+1}\,
            S\bigl(\mathbf{x}_{0:K}^{(i+1)},\mathbf{\Omega}_{1:K}^{(i+1)}\bigr).
        \]

        \State \textbf{M-step:}
        \State \quad Update $\bm{\theta}^{(i+1)}$ in closed form from $\widehat{S}^{(i+1)}$
       
    \EndFor

    \State \textbf{Output:} Estimated parameters $\widehat{\bm{\theta}}$ and 
state trajectory $\widehat{\mathbf{x}}_{0:K}$.
\end{algorithmic}
\end{algorithm}

\section{Experiments}
\noindent\emph{A. Synthetic case study: rotating ring source under RFI contamination}

This experiment evaluates the ability of the proposed robust SAEM algorithm to jointly
estimate a sequence of latent images and the model parameters in the context of
radio interferometric imaging. Radio interferometers measure incomplete and noisy
Fourier samples (visibilities) of the sky brightness distribution \cite{thompson},
and recovering the underlying images constitutes an ill-posed inverse problem that is
particularly sensitive to non-Gaussian noise such as radio-frequency interference
(RFI).

We generate a sequence of $K=10$ images of size $n = 64\times 64$ pixels, modeling a
ring-shaped astronomical source undergoing a cumulative rotation over time. The latent
dynamics follow
\[
    \mathbf{x}_k
    = \mathbf{F}\,\mathbf{x}_{k-1} + \mathbf{w}_k,
    \qquad
    \mathbf{w}_k \sim \mathcal{N}\!\big(\mathbf{0},\,\alpha\,\mathbf{I}_n\big),
\]
with $\alpha = 10^{-4}$ and where
$\mathbf{F} = \mathbf{S} \circ \mathbf{R}(\phi_k)$ is the composition of a 2D rotation by an angle
$\phi_k$ with the bilinear interpolation operator $\mathbf{S}$ acting on the discrete
image grid. 
Visibility measurements $\mathbf{y}_{1:K}$ are generated using the observation model
\eqref{eq:observation_model}, where
$\mathbf{H} \in \mathbb{C}^{m \times n}$ denotes the interferometric measurement
operator associated with incomplete Fourier sampling. Its analytical expression is
\begin{equation}
    \mathbf{H}
    =
    \exp\!\left(
        \frac{-j\,2\pi}{\lambda }\,
        \bm{\Delta}_{\mathbf{r}} \bm{L}^{T}
    \right),
    \label{FO}
\end{equation}
where $\lambda$ is the observing wavelength,
$\bm{\Delta}_{\mathbf{r}}$ is the $m\times 2$ baseline matrix containing the 2D
Cartesian antenna separations $(\mathbf{r}_i - \mathbf{r}_j)$, and $\bm{L}$ is the
$n\times 2$ matrix of direction cosines $(l,m)$ indexing the sky grid. To emulate an interferometric acquisition, we adopt the \emph{uv}-coverage of the VLA 
(27 antennas) at $3.8$~GHz, yielding $m=351$ visibilities per time step. The synthetic visibilities are corrupted following the proposed approach in \cite{leshem}, with a contamination rate of $15\%$ 
producing heavy-tailed deviations from the
ideal Gaussian noise assumption.

\textit{Initialization: } The initial state estimate $\bm{\mu}_0$ is set to the \emph{dirty image}, i.e., the inverse discrete Fourier transform of the measured visibilities, and the initial covariance is empirically chosen as $\mathbf{\Sigma}_0 = 10^{-3}\mathbf{I}_n$. For simulation convenience, the degrees of freedom parameter is fixed to $\nu = 2.5$, an empirically chosen value
that performs well in this setting. In principle, $\nu$ could also be estimated from
the data using, for example, one-dimensional line search or the rational approximation
method of \cite{nuref}. The  step-size sequence $(\gamma_i)$ is chosen according to $\gamma_i = 1/i$.

We compare the proposed robust SAEM method with two reference approaches: (i) a standard
Gaussian EM algorithm \cite{shum} designed for linear-Gaussian state-space estimation
under a strictly Gaussian noise assumption, and (ii) an oracle RTS smoother with access to the
true model parameters, which represents a theoretical upper bound in the Gaussian
regime.

\vspace{0.4em}

\noindent\emph{B. Reconstruction results}

Figure~\ref{fig:rotation_reconstruction} displays the reconstructed images at four
representative time steps. The robust SAEM method faithfully recovers the morphology of
the rotating ring source and accurately tracks its rotational motion despite significant
RFI contamination. As time increases, the reconstructions also tend to sharpen, since
the SAEM algorithm exploits the dynamical model to accumulate information across the 
sequence. Quantitative results in Table~\ref{tab:metrics}, based on RMSE, PSNR, and SSIM, confirm
these observations. 

The proposed method substantially outperforms both the Gaussian EM algorithm and the
oracle RTS smoother. This highlights that, in the presence of RFI-induced
heavy-tailed noise, robustness is more crucial than having exact knowledge of the true
model parameters. Overall, this experiment demonstrates the dual ability of the robust
SAEM framework to infer a partially unknown dynamical model while effectively mitigating
non-Gaussian noise.
\begin{figure}
\centering
\includegraphics[width=0.75\columnwidth]{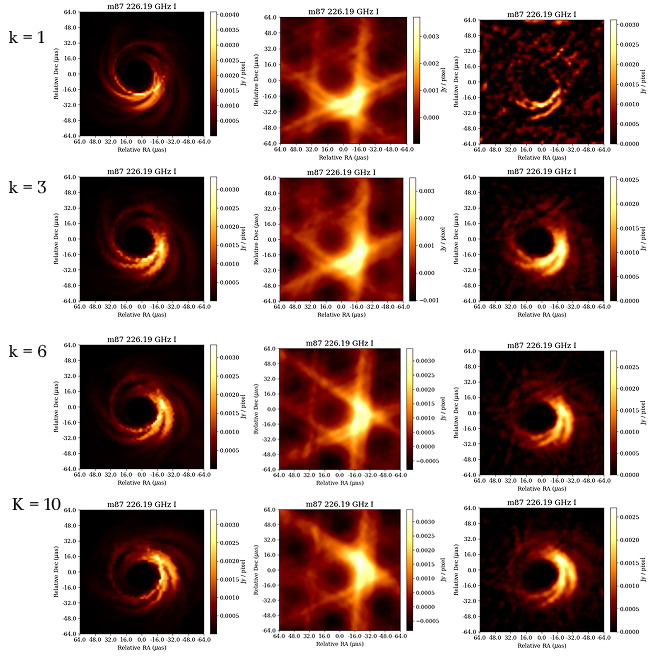}
\caption{Dynamic reconstruction using the proposed SAEM algorithm from simulated VLA visibilities. Comparison between ground-truth images (left) and SAEM estimates (right) at time steps $k \in \{1, 3, 6, 10\}$ (top to bottom).}
\label{fig:rotation_reconstruction}
\end{figure}

\begin{table}[t]
    \centering
    \caption{Comparison of reconstruction performance (sequence-averaged metrics)}
    \label{tab:metrics}
    \begin{tabular}{lccc}
        \toprule
        Method & MSE $\downarrow$ & SSIM $\uparrow$ & PSNR $\uparrow$  \\
        \midrule
        Gaussian EM & 0.00242 & 0.7635 & 26.905 dB  \\
        RTS smoother & 0.00147 & 0.8672 & 31.6438 dB \\
        \textbf{Proposed SAEM} & \textbf{0.00098} & \textbf{0.8935} & \textbf{34.1436 dB} \\
        \bottomrule
    \end{tabular}
\end{table}
\section{Conclusion}

We presented a robust SAEM framework for inference in heavy–tailed linear
state--space models with complex-valued observations. By exploiting the
conditionally Gaussian structure of the compound-Gaussian model, the method enables
closed-form Gibbs updates for both the latent states and the texture variables, and
relies on a complex-to-real FFBS sampler for efficient state simulation.
In a radio interferometric imaging scenario affected by RFI, the proposed approach
significantly outperforms a Gaussian EM algorithm and even an oracle RTS smoother.
These results demonstrate the clear benefit of heavy-tailed modeling for reconstruction
under non-Gaussian interference, and highlight the relevance of robust EM-type methods
for imaging in contaminated measurement regimes.
{

}
\end{document}